# New optimization scheme to obtain interaction potentials for oxide glasses

Siddharth Sundararaman[a,b], Liping Huang[a,†], Simona Ispas[b] and Walter Kob[b]

[a]Department of Materials Science and Engineering, Rensselaer Polytechnic Institute, Troy, NY 12180, United States

[b]Laboratoire Charles Coulomb (L2C), Univ Montpellier, CNRS, Montpellier, France

**Abstract**

We propose a new scheme to parameterize effective potentials that can be used to simulate atomic systems such as oxide glasses. As input data for the optimization, we use the radial distribution functions of the liquid and the vibrational density of state of the glass, both obtained from *ab-initio* simulations, as well as experimental data on the pressure dependence of the density of the glass. For the case of silica, we find that this new scheme facilitates finding pair potentials that are significantly more accurate than previous ones even if the functional form is the same, thus demonstrating that even simple two-body potentials can be superior to more complex three-body potentials. We have tested the new potential by calculating the pressure dependence of the elastic moduli and find a good agreement with the corresponding experimental data.

## 1. Introduction

Atomistic simulations like molecular dynamics (MD) allow determination of various properties of materials and hence to get valuable insight into their structure-property relations.[1–11] The key element of such studies is the accuracy of the interaction potential since all predicted properties of the material will depend to some extent on this potential, and sometimes this dependence can be surprisingly strong.[12] As a consequence, there is a long history of potential development for technologically important materials such as oxide glasses.[11,13–24] These studies clearly demonstrate the various obstacles one faces for finding a reliable effective potential for a given material: One has to decide on the functional form of the potential, on the reference data set (which experimental or *ab-initio* data), i.e., which quantities one attempts to reproduce, the thermodynamic state point(s) (temperature, pressure) at which the potential is aimed to be reliable, etc.[25] Even if all these decisions have been made one still faces the problem to find a set of parameters that minimize the cost function, i.e., makes the deviations to the reference data as small as possible. Experience shows that the cost function is a very complex function of the parameters of the potential (often way more than ten) and has many local minima. Hence finding the optimal set of parameters is a very difficult numerical task and so far no good solution exists.

---

[†]Corresponding author. E-mail address: huangL5@rpi.edu (L. Huang).

Most past efforts for the development of effective potentials have focused on the choice of the functional form. The first potentials were relatively simple because this makes the numerical simulations less costly and also limits the number of parameters that have to be determined.[13,19,26–28] However, these simple potentials often predict material properties that do not agree very well with the experimental findings and therefore people started to use more complex functional forms that usually contain dozens of parameters.[29] Although these complex potentials can sometimes give more reliable predictions for the properties of materials, it is usually cumbersome to obtain their parameters. Furthermore, it is often not clear whether the chosen parameter set is really the optimal one, since an exhaustive search in high dimensional parameter space is too costly, and last but not least the resulting potentials are often computationally expensive. It is therefore useful to check whether it is possible to use a simple functional form for the potential and improving its reliability by choosing a better set of reference data and a more efficient search algorithm to determine the best set of parameters. The goal of the present paper is to show that it is indeed possible to make a significant improvement of the quality of a simple potential by choosing a more optimum set of reference data.

The system we consider is amorphous silica, an open network glass-former that has a multitude of applications in science and technology. Because of this widespread use, there is a large body of experimental data that can be used to check the accuracy of the predictions made by simulations.[30,31] Therefore amorphous $SiO_2$ is an excellent system to check the efficiency of methods to develop effective potentials.

For the case of oxide glasses, one popular choice for the short range part of the potentials is the functional form proposed by Buckingham.[13,16,19,22,26,32] Van Beest et al. used, for example, this type of potential, supplemented with Coulombic interactions with fixed partial ionic charges, and optimized the parameters to reproduce first principles data of a $H_4SiO_4$ cluster and bulk properties of $α$-quartz.[13,22] The resulting BKS potential is able to predict a wide range of properties for silica glass and its crystalline counterparts with surprisingly good accuracy for such a simple potential form.[1,7,8,33,34] However, it was noted quite quickly that the BKS potential is not reliable regarding the vibrational density of states (VDOS) in the low and intermediate frequency range[1,2,16], that it predicted a much larger defect concentration on the surface as compared to *ab-initio* Car-Parrinello molecular dynamics (CPMD) calculations[6], and overestimated the elastic constants of silica glass.[35]

To fix some of these deficiencies, many-body effects like polarizability, fluctuating charges, or angular constraints, etc., were added to the two-body functional potential form but it was found that these additions do not necessarily give a more realistic description.[33] For example, the three-body potentials for silica glass developed by Vashishta et al.[20], Feuston and Garofilini[18], Huang and Kieffer[11,17], improve the elastic, structural and dielectric properties but still have many deficiencies. One such issue is that the accuracy of the potential quickly deteriorates once one probes temperatures or pressures that are significantly different from the ones for which the potential parameters have been optimized. This effect for example makes the simulations of nano-indentation difficult because of the presence of large pressure differences within the sample. The aim of the present work is therefore to develop an optimization scheme to parameterize computationally efficient pairwise potentials for oxide glasses to reliably predict various properties not just at ambient conditions but also over a large range of temperature and pressure. These potentials should be able to avoid some of the deficiencies of existing potentials like their inability to quench a glass sample with an acceptable density under normal conditions, to reproduce various features of VDOS or accurately predict structure and elastic properties both at ambient conditions and under high pressures.

Properties that are generally predicted well in MD simulations, such as bond length, cohesive energy, density, etc., depend mainly on the position and depth of the local potential well. (Note that this well is in the *effective* potential seen by a particle, i.e., it is the sum of many terms in the *bare* interaction potential.) The inability of potentials to accurately reproduce the vibrational and elastic properties suggests that it is the *shape* of this potential minimum that needs to be improved since these quantities also depend on this curvature. In other words, the VDOS depends on the different features of the interaction potential as a whole and hence its addition should allow to maintain the location and depth of the minimum of the interatomic interaction potential while substantially improving the shape as well as the 3-D atomic structure. The same is true for adding information of the system at elevated pressures/temperatures since under such conditions parts of the potential energy surface can be probed that are inaccessible at ambient conditions.

Our approach is hence to use as reference data for the fitting the results from accurate first principles calculations and explicitly include into the cost function the radial distribution functions (RDFs) of the equilibrium liquid as well as the VDOS and density of the glass. We use the RDFs of the liquid at multiple temperatures and the density at room temperature to take into account the temperature dependence of the structure. Density at various pressures at room temperature was included in the cost function to improve the response of glass to pressure. Note that all of these properties can be easily obtained using first principles calculations with reasonable system sizes, i.e., a few hundred atoms. This is especially important for extending the fitting approach to multi-component glasses, for which there is often little or no experimental data available for parameterizing classical potentials.

In the present work, the optimization scheme was applied to the Buckingham functional form with the Coulombic interactions described using the Wolf summation method[36] to further improve the efficiency of the calculations. Similar studies using other functional forms like the Morse form also showed improvements over existing potentials of the same functional form like Pedone potential[15], especially in quench densities and VDOS. In this work we chose the Buckingham functional form as in general it gave better results than the Morse functional form during optimization runs. In the following we will show how the choice of the quantities used in the fitting procedure affects the quality of the resulting potential and that a simple two-body potential can give a surprisingly good description of amorphous $SiO_2$.

The organization of the paper is as follows: In the next section we will present the details of the fitting procedure and discuss the quantities we have used in the cost function. In Sec. 3 we will compare the results obtained from the new potentials with the ones from other popular potentials for SiO₂ as well as experimental data. In particular we will demonstrate that our new potentials are significantly more accurate than the ones that can be found in the literature. In the final section we summarize our results and draw some conclusions on this new fitting approach.

## 2. Simulation methods

In this section we give the details of the minimization procedure as well as the quantities that are taken into account in the cost function. Furthermore we also describe how we determined the properties that we later use to check the accuracy of the potential.

### 2.1 Cost function for the Optimization

The functional form that we used for the interaction potential is the one proposed by Buckingham[37], i.e.,

$$V^{Buck}(r_{\alpha\beta}) = \frac{q_\alpha q_\beta e^2}{r_{\alpha\beta}} + A_{\alpha\beta} exp(-B_{\alpha\beta} r_{\alpha\beta}) - \frac{C_{\alpha\beta}}{r_{\alpha\beta}^6} + \frac{D_{\alpha\beta}}{r_{\alpha\beta}^{24}} \quad (1)$$

where $r_{\alpha\beta}$ is the distance between a pair of particles of species $\alpha$ and $\beta$ ($\alpha, \beta$ = Si, O), $e$ is the elementary charge and $\frac{D_{\alpha\beta}}{r_{\alpha\beta}^{24}}$ is a repulsive term added to avoid the divergence of the potential due to the van der Waals term at very small distances, which can cause problems for simulations at high temperatures. We have found that the quality of the potential does not improve significantly if one uses an attraction term for the cation-cation pair and therefore we set $C_{SiSi} = 0$. The rest of the parameters are grouped into the vector $\phi_k (k = 1,..., N_{param})$, where $N_{param}$ is the number of fit parameters considered in the optimization. Note that only a single charge term appears in the above vector since charge neutrality of the system requires $q_O = \frac{-q_{Si}}{2}$. The long range Coulombic interactions, $V^W(r_{\alpha\beta})$, are evaluated using the Wolf truncation method,[36,38]

$$V^W(r_{\alpha\beta}) = q_\alpha q_\beta \left[\frac{1}{r_{\alpha\beta}} - \frac{1}{r_{cut}^W} + \frac{(r_{\alpha\beta} - r_{cut}^W)}{(r_{cut}^W)^2}\right] \quad (2)$$

where the long range cutoff is set to $r_{cut}^W$ =10 Å. The Wolf method for evaluating the Coulombic interaction was found to be orders of magnitude faster than the Ewald method[39], especially for systems of the order of 10000 atoms or larger, in agreement with a previous study by Carré et al.[38] One of the main goals of this paper was to optimize an efficient pair potential that is able to reproduce properties like elastic properties and their response to thermodynamic stimuli better.

Hence having a potential that has a finite interaction range will enable us to perform large scale mechanical tests like nano-indentation reliably in a reasonable amount of time. To prevent discontinuities in the potential or the force we have multiplied the above potentials with the function

$$G_{r_{cut}}(r_{ij}) = \exp\left(-\frac{\gamma^2}{(r_{ij} - r_{cut})^2}\right) \qquad (3)$$

where $\gamma = 0.2$ Å defines the width of the smoothing function and $r_{cut}$ is the cutoff distance.[40] For integrating the equation of motion we used a timestep of 1.6 fs and, when necessary, a Nosé-Hoover thermostat[41,42] and Parrinello-Rahman barostat[43] were employed for maintaining constant temperature and pressure, respectively. All the molecular dynamics simulations were carried out using the large-scale atomic/molecular massively parallel simulator (LAMMPS).[44]

Carré et al., put forward the idea to use structural information of high temperature liquid to determine the parameters of the effective potential.[16,40] In practice they used the partial RDFs as obtained from *ab-initio* simulations and a Buckingham potential as the functional form. The resulting potential (CHIK potential) from that work gave better VDOS feature at intermediate frequencies than that predicted by the BKS potential, but certain features of the experimental VDOS were not reproduced in a satisfactory manner.[16] Very similar to the BKS potential, the CHIK potential could not reproduce very well the elastic moduli at ambient conditions.[34] In the present work, the approach of Carré et al. is extended to include in the cost function the RDFs of the liquid at multiple temperatures, the VDOS of glass at 0 K, and the density of glass at 300 K at various pressures. The resulting cost function that measures the quality of the effective potential is therefore given by

$$\chi^2(\phi) = w_1 \sum_{T=T_1,T_2} \int_0^{r_{N_{RDF}}} \sum_{\alpha,\beta} (rg_{\alpha\beta}^{calc,T}(r|\phi) - rg_{\alpha\beta}^{ref,T}(r))^2 dr + \\ w_2 \int_0^{\nu_{N_{VDOS}}} \sum_{\alpha} (f_\alpha^{calc}(\nu|\phi) - f_\alpha^{ref}(\nu))^2 d\nu + w_3 \sum_{n=1}^{N_{Pres}} (\rho^{calc}(P_n|\phi) - \rho^{ref}(P_n))^2 \qquad (4)$$

where $\phi$ is the current parameter set, $\alpha, \beta$ are the different species, $w_1, w_2, w_3$ are the weights for each contribution, $rg_{\alpha\beta}(r)$ is the RDF weighted by the distance $r$ up to a maximum distance of $r_{N_{RDF}} = 6$ Å (the reason for using the factor $r$ can be found in Carré et al. [16]), $f_\alpha(\nu)$ is the partial VDOS at frequency $\nu$ at 0 K up to a maximum frequency of $\nu_{N_{VDOS}} = 1400\,\text{cm}^{-1}$, $\rho$ is the density at 300 K for various pressures up to a maximum pressure of 16 GPa, and $T_1, T_2$ are the two temperatures (3600 K and 4800 K) at which the RDF was evaluated. The superscript 'ref' refers to the first principles or experimental reference data towards which the optimization was carried out, and superscript 'calc' refers to the calculated properties using the current parameter set. Each part of the cost function was normalized so that their magnitudes are comparable. The weight coefficients were then chosen to achieve the best trade-off between the various properties in the statistical sampling we have done and in general depend on the system under consideration. For the present case, we used $w_{RDF} = 0.25$ for both temperatures, $w_{VDOS} = 0.35$ and $w_{Density} = 0.15$.

For the calculation of the RDF, a sample of 1200 atoms was first equilibrated in the NPT ensemble (constant number of atoms, pressure and temperature) at the reference data pressure (see below) for 20 ps. This was followed by a production run in the NVT ensemble (constant number of atoms, volume and temperature, with the volume given by the average volume over the NPT runs) for 40 ps to measure the RDF. Samples at low temperatures are required for the calculation of the VDOS. This necessitates quenching configurations at high temperature to 0 K, a procedure that is computationally very costly during optimization. To avoid these quenching simulations, we have thus taken the first principles reference structure in the glass state, i.e., a sample with 384 atoms, and relaxed it to its nearest local minimum using the effective potential. Subsequently we determined the corresponding dynamical matrix, see section 3.2 for details, and then the VDOS. To test the accuracy of this shortcut we have compared for the optimal effective potential the so-obtained VDOS with the VDOS as determined from a glass sample that was quenched from high temperatures. Since no significant difference was found between the two densities of state we can conclude that our approach is sufficiently accurate.

The minimization was performed using the Levenberg-Marquardt algorithm[45,46] and the numerical derivatives have been calculated using a finite difference method. The parameter set corresponding to the minimum of the cost function is given in Table 1. This parameter set, optimized for the Buckingham potential with a short-range cutoff (for non-Coulombic interactions) of 8 Å, will be referred to as "SHIK-1" in the rest of the paper. As we will see below, it is this set of parameters that can be considered to be the best for amorphous $SiO_2$.

**Table 1:** Parameters for SHIK-1.

| i-j | $A_{ij}$ (eV) | $B_{ij}$ (Å$^{-1}$) | $C_{ij}$ (eV·Å$^6$) | $D_{ij}$ (eV·Å$^{24}$) |
|---|---|---|---|---|
| O-O | 1120.5 | 2.893 | 26.1 | 16800 |
| O-Si | 23107.8 | 5.098 | 139.7 | 66.0 |
| Si-Si | 2797.9 | 4.407 | 0.0 | 3423204.0 |

$q_{Si}$ = 1.7755e, $r_{cut}^{SR}$ =8 Å, $r_{cut}^{LR}$ =10 Å

To test the influence of the cost function on the quality of the potential, we have also considered the case that the cost function only includes the density at ambient pressure, i.e., no pressure dependence of the density was taken into account. The resulting set of parameters will be referred to as "SHIK-2", and it is given in Table 2. Note that the short range cutoff used for this potential was 5.5 Å, and below we will see that this cutoff plays an important role especially in the resistance of the structure to compression. Parameters for the BKS potential are given in Table 3 for reference. The short range cutoff for this potential was chosen to be 5.5 Å to get the glass density close to experimental data.[1]

**Table 2:** Parameters for SHIK-2.

| i-j | $A_{ij}$ (eV) | $B_{ij}$ (Å$^{-1}$) | $C_{ij}$ (eV·Å$^6$) | $D_{ij}$ (eV·Å$^{24}$) |
| --- | --- | --- | --- | --- |
| O-O | 1085 | 2.95 | 28.8 | 16800 |
| O-Si | 22707 | 5.135 | 130.6 | 66.0 |
| Si-Si | 3876.9 | 3.93 | 0.0 | 3423204.0 |

$q_{si}$= 1.74e, $r_{cut}^{SR}$ =5.5 Å, $r_{cut}^{LR}$ =10 Å

**Table 3:** Parameters for BKS.

| i-j | $A_{ij}$ (eV) | $B_{ij}$ (Å$^{-1}$) | $C_{ij}$ (eV·Å$^6$) |
| --- | --- | --- | --- |
| O-O | 1388.773 | 2.76 | 175.0 |
| O-Si | 18003.757 | 4.87318 | 133.538 |
| Si-Si | 0.0 | 0.0 | 0.0 |

$q_{si}$= 2.4e, $r_{cut}^{SR}$ =5.5 Å, $r_{cut}^{LR}$ =10 Å

## 2.2 Reference data generation

First principles MD simulations were carried out using the Vienna *ab-initio* package (VASP).[47,48] The electronic structure was described by means of the Kohn–Sham (KS) formulation of the density functional theory[49] using the generalized gradient approximation (GGA) and the PBEsol functional.[50,51] The KS orbitals were expanded in a plane-wave basis at the Γ point of the supercell, including components with energies up to 600 eV, with the electron-ion interaction described within the projector-augmented-wave formalism.[52,53] For the solution of the KS equations, we used the residual minimization method-direct inversion in the iterative space[47], and the electronic convergence criterion was fixed at 5 × 10$^{-7}$ eV. The choice of the above approximations and parameters were motivated by previous *ab-initio* studies of liquid and glass of a sodium borosilicate system.[54,55]

The system consisted of 384 atoms, i.e., 128 SiO$_2$ units, confined in a cubic box with periodic boundary conditions. The edge length of the box was fixed to 17.96 Å, which corresponds to the experimental density of 2.2 g/cm$^3$ at ambient conditions for silica glass.[30] First principles MD simulations were done in the NVT ensemble at 4800 K and 3600 K, by using a Nosé thermostat[41] to control the temperature. The starting points for the configurations at both temperatures were taken from equilibrium classical MD simulations using the CHIK potential.[16]

At 4800 K, the total length of the run was 2.2 ps, and we discarded the first 0.5 ps before we started the measurement of the static properties, notably the radial distribution functions $g_{\alpha\beta}(r)$ needed as input for the cost function in Eq. (4). At 3600 K, we equilibrated the sample for 1.5 ps, and we used the following 10.7 ps of the run to measure the structural properties to be used for the potential optimization.

In order to prepare a glass sample and compute its vibrational properties, we chose not to quench the *ab-initio* liquid, and instead generate a glass sample within a classical MD approach using the CHIK potential. This sample contained 384 atoms, confined in cubic box of 17.96 Å, i.e., corresponding to the experimental density of silica glass at ambient conditions and was equilibrated for 500 ps at 4000 K and then quenched to 300 K with a cooling rate of about 1 K/ps in the NVT ensemble. Using the so-obtained configuration as the starting point, we performed an *ab-initio* NVT run at 300 K for 2.3 ps, followed by a short NVE run of 0.6 ps. The final configuration was relaxed to 0 K and then we determined its dynamical matrix using a finite difference scheme with atomic displacements of about 2.6 x 10$^{-3}$ Å. We adopted this combined classical and *ab-initio* procedure as it allows saving a large amount of CPU time and has been shown to reproduce VDOS in very good agreement with experimental data.[56,57] Relaxing glass under high pressures using first principles methods involves significant computation cost, hence we used readily available experimental data. The pressure dependence of the density was obtained from experiments at room temperature by S. P. Jaccani at RPI using optical microscopy with a diamond anvil cell in the pressure range from ambient conditions to 16 GPa (private communication).

### 2.3 Glass preparation
To determine the accuracy of the effective potential in predicting the properties of SiO$_2$ glass we used the melt-quench method to prepare glass samples. We first equilibrated samples of 10368 atoms at a density of 2.2 g/cm$^3$ at 4000 K in the NVT ensemble for about 1 ns and then quenched them in the NPT ensemble to 300 K in steps of 50 K at a nominal cooling rate of 1 K/ps while maintaining zero pressure. Based on previous studies in literature[1,58,59], this cooling rate is in the region where the cooling rate effect on structure and properties of silica glass is small. Quenched samples were then annealed at 300 K in the NPT ensemble for another 100 ps.

Samples using the BKS potential[13] were prepared by first relaxing a 10368 atoms system at a density of 2.2 g/cm$^3$ at 6000 K for 200 ps in an NVT ensemble, which were then brought down to 4600 K in the NVT ensemble and quenched in the NPT ensemble to 300 K at a cooling rate of 5 K/ps maintaining zero pressure, followed by relaxation at 300 K for another 100 ps in the NPT ensemble. The long range Coulombic interactions were calculated using the Ewald summation method[39] using a cutoff of 10 Å with a K-space accuracy of $10^{-4}$. The short range interactions were cut off at 5.5 Å to get a reasonable quench density.[3] To improve the statistics of the results, 4 independent samples were used in each case.

## 3. Results and Discussion

In this section we test and discuss the ability of the new potential to reproduce the *ab-initio* and experimental data for amorphous silica in the liquid and glass state. For the sake of comparison we also include some data as obtained from the BKS potential.

### 3.1 Properties of liquid

One of the simplest macroscopic quantities of a liquid is its density ρ. Figure 1 shows the temperature-dependence of ρ at zero pressure when the system is cooled from high temperatures to 300 K. The curve for the SHIK-2 potential show at around 3250 K a pronounced maximum, the one for the BKS potential has a maximum at around 4500 K,[1] and the data for the SHIK-1 potential shows a weak maximum around 3000 K. These findings can be compared with the experimental data[60] measured as a function of the fictive temperature ($T_f$), which shows a weak density maximum at $T_f$ around 1850 K. Thus we can conclude that the SHIK-1 and SHIK-2 potentials are more reliable than the BKS potential with respect to the location of this density anomaly. For the cooling rate (1 K/ps) used here, the kinetic glass transition temperature calculated from the change in slope of the curves in Fig. 1 is around 2000 K for the new potentials and 3000 K for the BKS potential. For temperatures below this glass transition, we find that the SHIK-2 and BKS predict only a weak variation in density, around 0.5%, in agreement with experiments[30], whereas the SHIK-1 potential predicts a slightly larger change in this temperature range (around 1.25%).

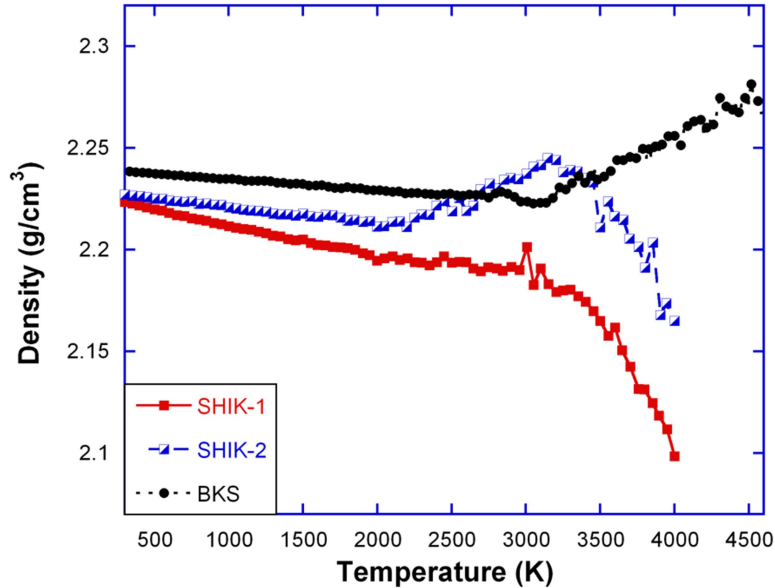

**Fig. 1** Density as a function of temperature for silica during NPT quenching from SHIK-1 and SHIK-2 and in comparison to the BKS potential.

The RDFs discussed in Sec. 2 give information regarding the relative arrangements of the particles. In Fig. 2 we show the partial RDFs of the silica liquid at 3600 K from SHIK-1 and SHIK-2 as well as the ones predicted by the BKS potential and the reference *ab-initio* data. We recognize that the new potentials show a much better agreement with the reference data as compared to the BKS potential. This is of course not that surprising since these partial RDFs are included in the cost function used to determine the optimal set of parameters. It is, however, remarkable that for this observable there is basically no difference between the data for SHIK-1 and SHIK-2 which shows that the RDFs are not affected by the difference in the used cost function. Figure 3 shows the bond angle distributions (BADs) of the silica liquid at 3600 K. Although these BADs were not explicitly included in the cost function, the new potentials give a significantly better agreement with the reference data as compared to the BKS potential. It is important to note here that even though the BADs are not completely independent of the RDFs, they are not entirely determined by them either. Also here we can see that the curves for SHIK-1 are very similar to the ones for SHIK-2, showing that the pressure dependence of the density in the cost function is not very relevant for this observable. Similar observations were made in RDFs and BADs at 4800 K (data not shown).

Figure 3(b) shows that classical pair potentials with fixed charges tend to predict higher Si-O-Si angles as compared to first principles data. Certain 3-body potentials like the one by Huang and Kieffer also predicts higher Si-O-Si angles[61], while the ReaxFF potential predicts a much lower Si-O-Si angle[62]. Thus, it seems not to be a limitation of the Buckingham form, but rather a general deficiency of empirical potentials, which may arise from the difficulty in properly describing the lone pair electrons of oxygen atom to give the correct Si-O-Si angles.

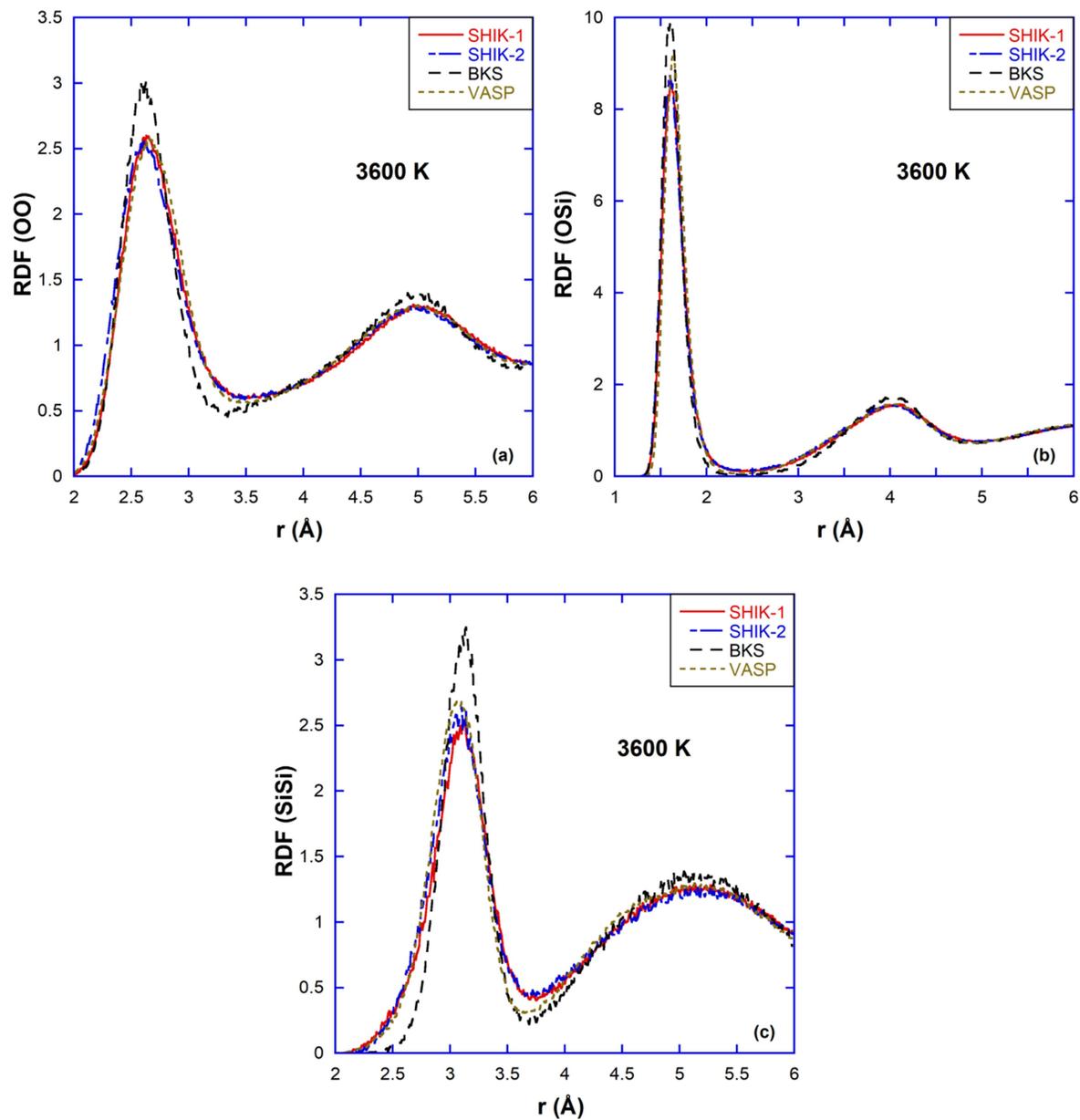

**Fig. 2** (a) O-O, (b) Si-O and (c) Si-Si RDF of silica liquid at 3600 K from SHIK-1, SHIK-2, and BKS potentials as compared to the reference VASP data.

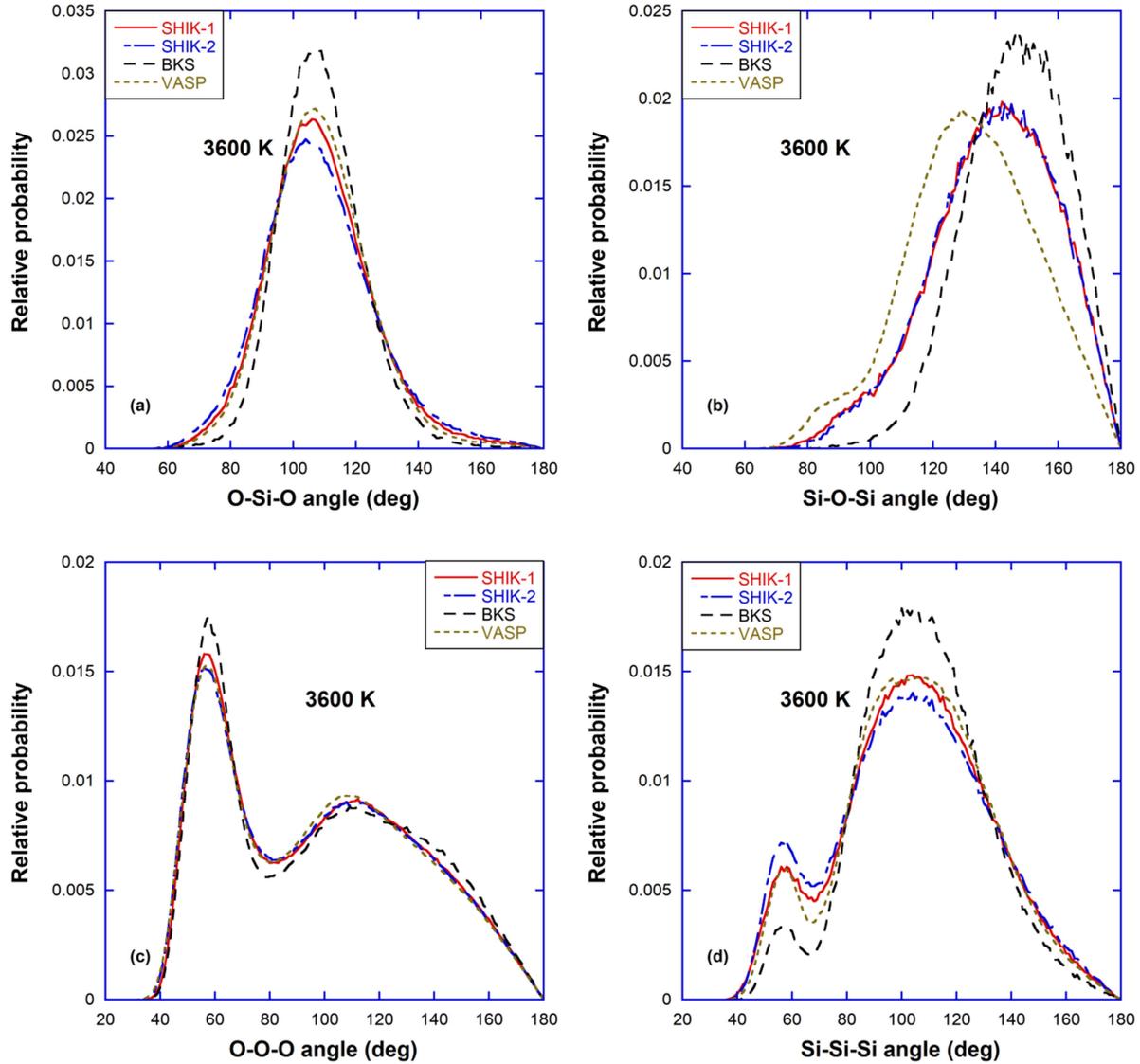

**Fig. 3** Bond angle distributions of silica liquid at 3600 K from SHIK-1, SHIK-2, and BKS potentials as compared to the reference VASP data for (a) O-Si-O, (b) Si-O-Si, (c) O-O-O and (d) Si-Si-Si.

### 3.2 Properties of the glass at room temperature

The properties considered so far concerned the liquid state. In this subsection we will discuss the glass state, i.e., the properties of the system at room temperature.

The RDFs discussed in the previous section are not directly accessible in real experiments. However, their space Fourier transforms can be measured, e.g., in neutron scattering experiments. In Fig. 4 we show the neutron structure factors at 300 K as predicted by the various potentials at 0 and 8.2 GPa in panels (a) and (b), respectively, and compare them with experimental data.[63] The neutron structure factor, $S_N(q)$, can be calculated from the partial structure factors, $S_{\alpha\beta}(q)$, using the equation [64]

$$S_N(q) = \frac{N}{\sum_\alpha N_\alpha b_\alpha^2} \sum_{\alpha,\beta} b_\alpha b_\beta S_{\alpha\beta}(q) \qquad (5)$$

where $\alpha, \beta \in \{Si, O\}$, $b_\alpha$ is coherent neutron-scattering length, $N_\alpha$ is the number of atoms of species $\alpha$ and $N$ is the total number of atoms. We recall that the properties of a glass, and hence its structure, does depend on the cooling rate with which the sample was produced[64], but this dependence is relatively weak[1] and hence can be neglected for the moment. Panel (a) demonstrates that the different potentials predict structures that are all quite similar and close to the one found in experiments. This good agreement is of course not that surprising since the two body correlation function is part of the cost function for the optimization. If pressure is increased, as seen in panel (b), the four curves are very similar at intermediate and large $q$ but clear differences can be noted at smaller $q$ in that the intensity of the first peak (see inset) depends significantly on the considered potential. Since this peak characterizes the structure of the sample on the length scale of the tetrahedra, we can conclude that the way the sample reacts to pressure on an intermediate range length scale depends significantly on the potential considered, a result that will be confirmed below when we discuss the mechanical properties.

From panel (b) we also recognize that the potential SHIK-1 is able to reproduce very well the experimental data at 8.2 GPa, while the BKS potential shows a decent agreement, and the SHIK-2 potential predicts an intensity of the first peak that is significantly too low. So this trend shows the utility of including density data at higher pressure into the cost function as we did for the case of the SHIK-1 potential.

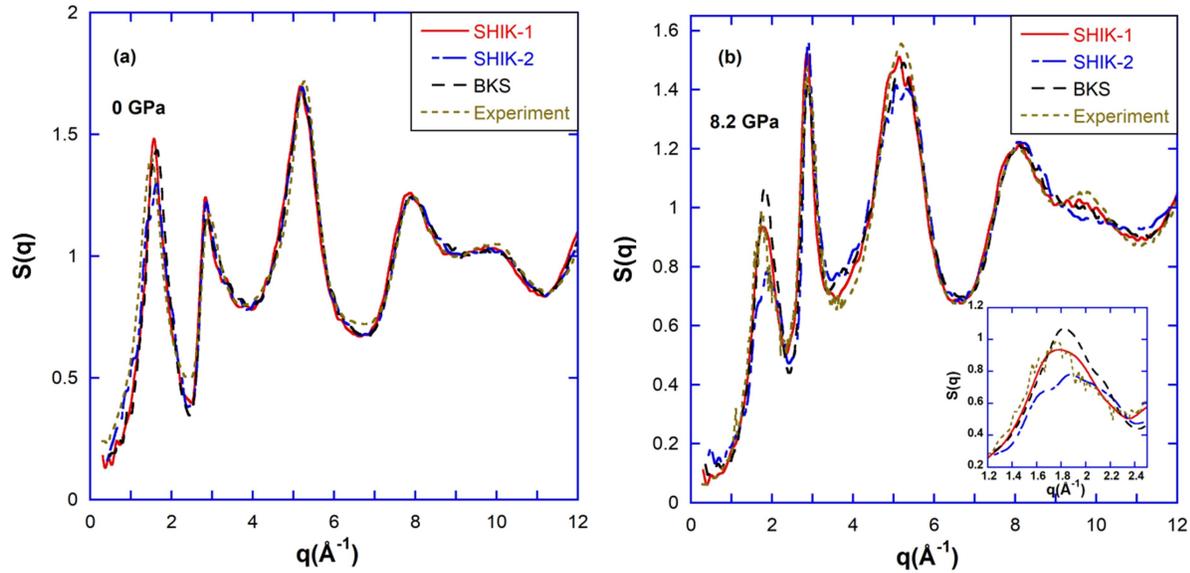

**Fig. 4** Neutron structure factor of silica glass at 300 K from the various potentials at (a) 0 GPa and (b) 8.2 GPa (the first peak shown in inset) in comparison to experiment[63].

The quick deterioration of the open network structure predicted by SHIK-2 with increasing pressures is also clearly seen in the distribution of the intra-tetrahedral angle (O-Si-O) and the intra-tetrahedral angle (Si-O-Si) shown in Fig. 5. Concerning the O-Si-O distribution at zero pressure, panel (a), one finds that at low pressure the distributions from the three potentials agree quite well with each other although differences can be noted in the intensity of the peaks. However, significant differences are observed if the pressure is increased to 8.2 GPa, as seen in panel (b). For the SHIK-2 potential the distribution has become much broader and shows a peak at around 80 degrees whereas the distributions for the two other potentials have just become a bit broader. Qualitatively the same behavior is seen for the distribution of the Si-O-Si angle in panels (c) and (d): At low pressures, the distributions from the classical potentials are very similar to each other with the SHIK-1 and SHIK-2 predicting the maximum at a slightly lower value as compared to BKS, while the structure relaxed with VASP predicts this peak at a lower value. Since experimental values for most probable Si-O-Si angle at ambient conditions reported in literature show a broad range from 142° to 160° (see in Fig 5(c)),[65] all of these predictions are acceptable. At higher pressure though, one finds again a marked difference in that the one for the SHIK-2 potential shows a strong peak at around 90 degrees, the typical signature for the presence of edge-sharing tetrahedra. This means that for this potential the structure has indeed started to transform from an open network like geometry to a more compact one.

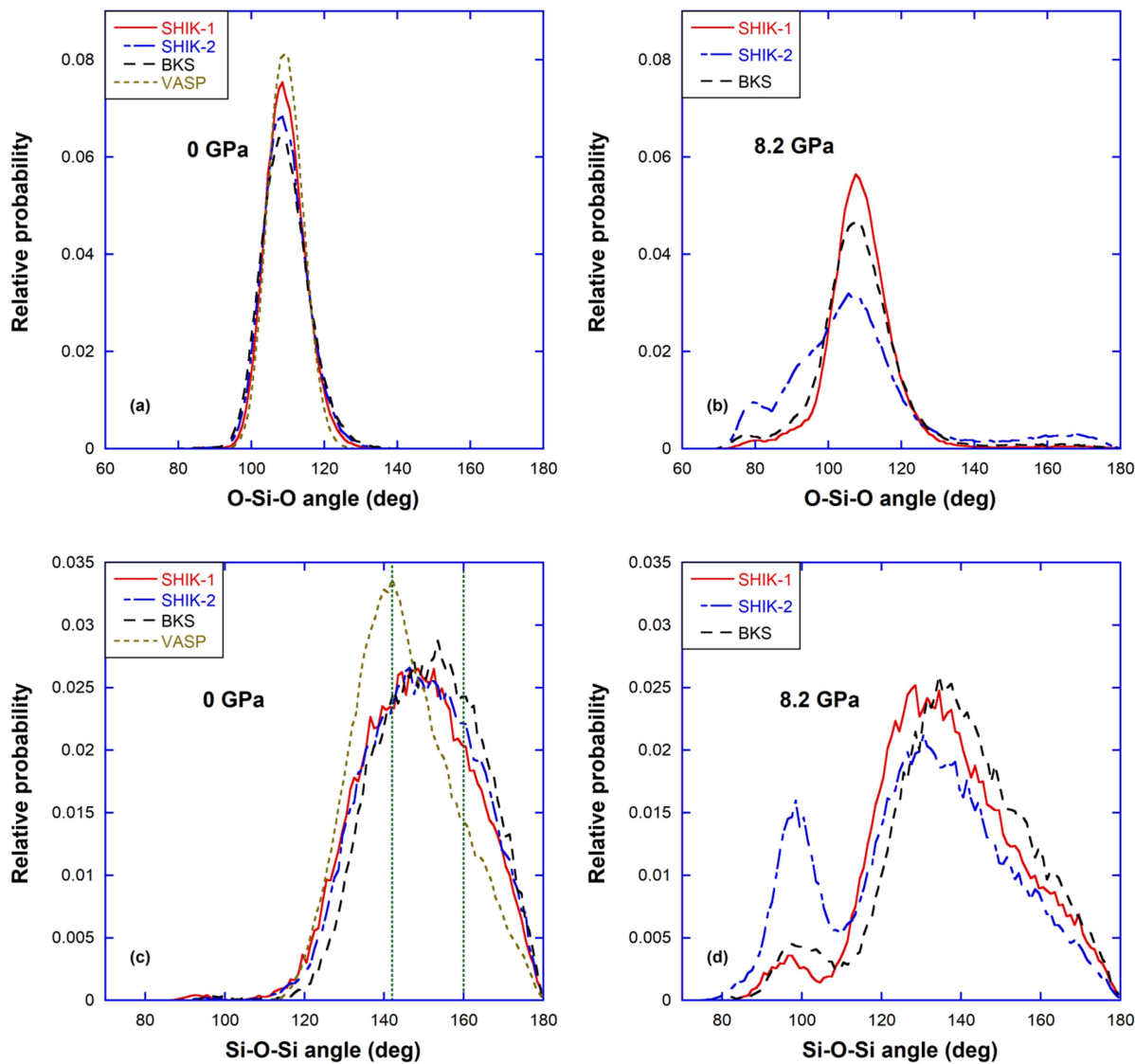

**Fig. 5** Bond angle distributions of silica glass at 300 K from SHIK-1, SHIK-2 and BKS potentials for (a) O-Si-O (0 GPa), (b) O-Si-O (8.2 GPa), (c) Si-O-Si (0 GPa) compared to VASP data and experimental data from various sources (shown as a range between the two vertical dashed lines)[65] and (d) Si-O-Si (8.2 GPa).

Although the static structure factor and the BAD give information on the local structure, it is not very informative regarding the particle arrangement on the intermediate length scales. Insight into this intermediate range order can be obtained from the distribution of the length of primitive rings, i.e., the shortest closed loop that includes a given Si atom and two of its nearest neighbor O atoms. To analyze how this intermediate range order changes with pressure, we have used the R.I.N.G.S code[66] to determine the primitive rings distribution.[67,68] The resulting ring statistics of the glass sample at 300 K from the various potentials at 0 and 8 GPa are shown in Fig. 6(a) and (b), respectively. For 0 GPa the ring size distribution from the three effective potentials do not differ much from each other and they are qualitatively very similar to the one from the VASP calculation even though the latter shows more 6-membered rings. In contrast to this, at 8 GPa a larger number of 2 and 3-membered rings are observed for SHIK-2 in Fig. 6(b), indicating that the open network structure starts to deteriorate. For this pressure the BKS potential also gives an increase of 2-membered rings, i.e., edge sharing tetrahedra, while SHIK-1 predicts only a very small concentration of these short rings. Thus we can conclude that SHIK-1 and BKS give structures that are more resistant to compression than SHIK-2.

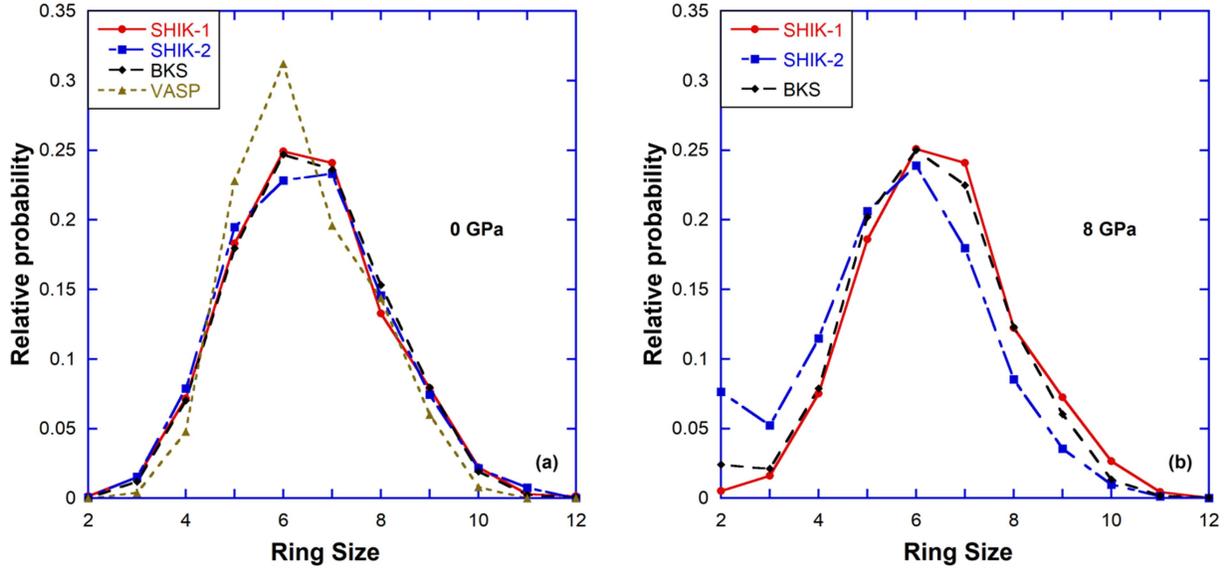

**Fig. 6** Ring statistics of silica glass at 300 K from the various potentials at (a) 0 GPa and (b) 8 GPa.

We now move on to the VDOS, i.e., a quantity that is related to the vibrational properties of the solid and that can be obtained from various spectroscopic measurement techniques.[69] The VDOS was determined by calculating the eigenvalues, $\omega^2$, of the dynamical matrix ($D_{i,\alpha,j,\beta}$).

$$D_{i,\alpha,j,\beta} = \frac{1}{\sqrt{m_i m_j}} \left( \frac{\partial^2 V(r)}{\partial r_{i,\alpha} \partial r_{j,\beta}} \right) \quad (6)$$

$$\omega^2 u_{i,\alpha} = \sum_{j,\beta} D_{i,\alpha,j,\beta} u_{j,\beta} \quad (7)$$

where $\alpha, \beta$ are the Cartesian coordinates $x$, $y$ and $z$, $i$, $j$ are the atoms in consideration, $u_{i,\alpha}$ is the displacement of the $\alpha$ coordinate of the $i^{th}$ particle from its equilibrium position for the normal vibrational mode with frequency $\omega$. The so-obtained discrete spectrum was convoluted with a Gaussian function with a half width at full maximum of 38 cm$^{-1}$ in order to smoothen the data.

In Fig. 7 we show the VDOS from the various potentials as compared to the one predicted by the *ab-initio* calculation. In agreement with previous results[1,2], we find that the BKS potential correctly predicts the splitting of the high frequency stretching modes with peak positions that are close to the ones obtained from the *ab-initio* approach or experiments.[56,57,69] However, at intermediate frequencies, the BKS potential is not able to give an accurate VDOS, e.g., it misses the marked peaks at around 400 cm$^{-1}$ and 750 cm$^{-1}$. This deficiency is much less severe for the SHIK-1 and SHIK-2 potentials since the corresponding VDOS does show these peaks, although less pronounced than the *ab-initio* data. The general features of the VDOS in the low and intermediate frequency range are better reproduced by the new potentials than the BKS potential. The splitting of the high frequency peak is a relatively delicate feature since the frequency of two types of modes needs to be correctly predicted. SHIK-1 is able to predict the splitting of the high frequency peak, although the two frequencies are closer to each other than those from the BKS potential and *ab-initio* data, while SHIK-2 shows a broad shoulder next to the main peak. If a lower charge was chosen for SHIK-1, the splitting of the high frequency peak becomes more obvious at the expense of deteriorating the VDOS in the low and intermediate frequency range. The charge for Si in Table 1 was chosen to give an overall better VDOS and other properties. Finally we note that the VDOS as predicted by these new simple pair potentials are more accurate than the ones obtained from other potentials that have a much more complicated functional forms.[20,61] Hence we conclude that better VDOS can be obtained even with simple functional forms of the potential.

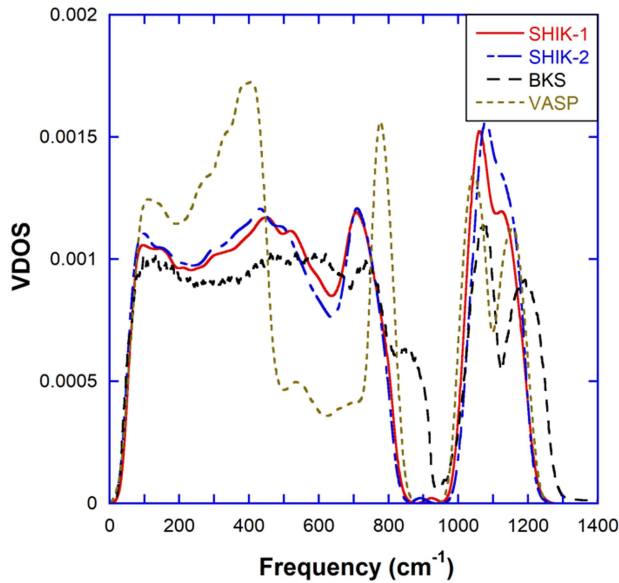

**Fig. 7** VDOS of silica glass at 0 K from SHIK-1, SHIK-2 and BKS potentials and compared to the reference VASP data.

The results presented above show that the structure at high and room temperature predicted by the two new potentials are very similar if the pressure is small. But this changes drastically if one applies pressure to these samples, as already seen in Figs. 4(b)-6(b), and hence this pressure dependence can be used to improve the reliability of the potentials. The variation of density and average Si coordination number with pressure for the various potentials are shown in Fig. 8. As expected, we find that the densification for the SHIK-2 potential is much more pronounced than the one found in experiment.[63] We also note that for this potential the slope of the density variation with pressure changes around 4 GPa, which can be related to the increasing coordination number of the Si atoms (5- or 6-fold) just above 4 GPa for this potential, as seen in Fig. 8(b). In contrast to this, the SHIK-1 and BKS potentials do not show such a drastic change in the pressure range of 0-8 GPa in Fig. 8(a), thus confirming that their corresponding network is still intact.

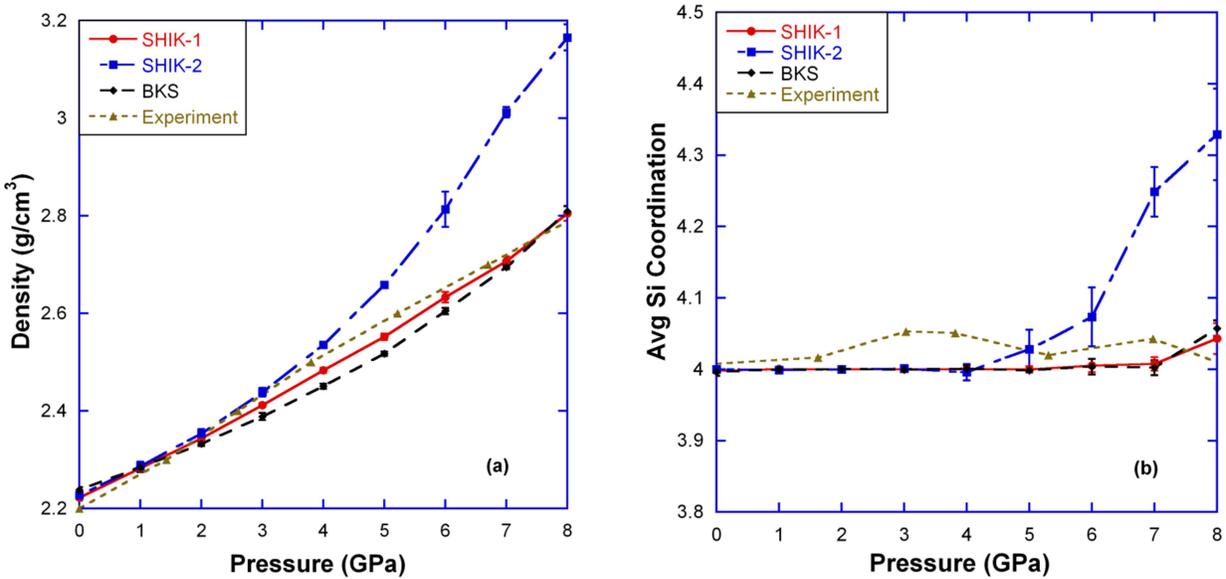

**Fig. 8** Variation of density (a) and average Si coordination (b) with pressure from the various potential in comparison to experiment.[63,70]

If a glass sample is densified beyond a certain threshold, it will compress inelastically, i.e., a release of the pressure will not bring it back anymore to the original density. Therefore another interesting test for the reliability of the potential is to check at which pressure this threshold is and how much the density changes if the sample is compressed beyond this value. Samples were compressed and subsequently decompressed at a nominal rate of 0.04 GPa/ps. The high compression rate is very similar to the high cooling rate used in MD, which is inevitably limited by the computational power. We tested different compression rates and chose one in a range where the results were not very sensitive to it. Figure 9 shows the density change in the recovered silica glass as a function of the maximum applied pressure during compression at 300 K for the SHIK-1 and BKS potentials as compared to the experimental data.[71] The permanent densification predicted by SHIK-1 is in good agreement with experiment and better than BKS especially at higher pressures.

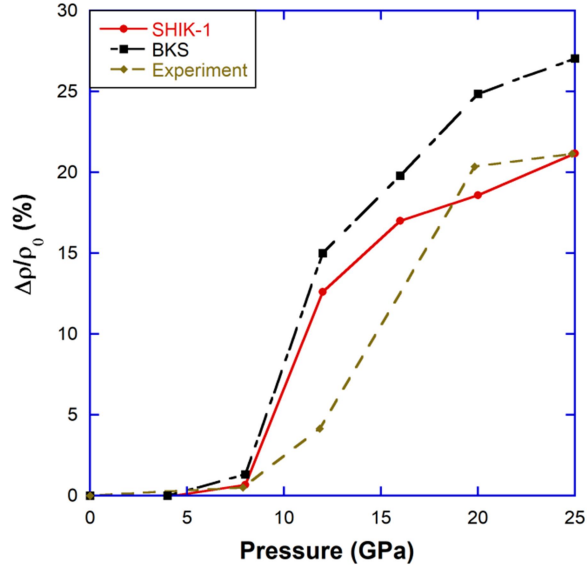

**Fig. 9** Density change in the recovered silica glass as a function of the maximum applied pressure during compression at 300 K from the SHIK-1 and BKS potentials in comparison to experiment[71].

Further quantities that are of great practical interest are the elastic moduli. The elastic moduli were measured by compressing and expanding the samples hydrostatically at a constant strain rate (1.25/ns) up to a volume change of 0.5% and measuring their stress response. The bulk modulus $B$ was then obtained from the relation

$$B = -V \frac{dP}{dV} = \frac{dP}{d \ln V} \qquad (8)$$

where $B$ is the bulk modulus, $V$ is the volume and $P$ is the pressure. For measuring the longitudinal modulus $C_{11}$, the samples were strained along the $x$, $y$ and $z$ axis in both the positive and negative direction at a constant strain rate to a linear strain of about 0.5% each, holding the cross-sectional area constant. The $C_{11}$ was then calculated from the slope of the stress-strain curve. The Young's modulus, $E$, and shear modulus, $G$, were obtained from $B$ and $C_{11}$ using the relations[72]

$$G = \frac{3}{4}(C_{11} - B) \quad (9)$$

$$E = \frac{9B(C_{11} - B)}{(3B + C_{11})} \quad (10)$$

The response of the elastic moduli to pressure was calculated by pressurizing samples in steps of 0.5 GPa, annealed at every pressure for 25 ps, before elastic moduli were calculated using the above procedures.

After having discussed above the anomaly in the density at high temperatures, we recall at this point another peculiar behavior of silica glass: the anomaly in the compressibility, i.e., the fact that with increasing pressure the system first softens until about 2-3 GPa[73–75] before becoming stiffer upon further compression. Although it is found that some potentials are able to reproduce this behavior, the compressibility minimum is usually found at too high pressures.[61,76,77] Figure 10 shows the pressure dependence of the different elastic moduli as predicted by the three effective potentials as well as the corresponding experimental data.[78] One recognizes from panel (a) that SHIK-1 and SHIK-2 both predict a bulk modulus of ~40 GPa at ambient conditions, close to the experimental value of 37 GPa. If pressure is increased, SHIK-2 predicts at 3.5 GPa a minimum in the bulk modulus with a value B=32 GPa and the corresponding values for the SHIK-1 are 3 GPa for the position of the minimum and 38 GPa for its depth, i.e., values that are quite close to the experimental ones. The BKS potential on the other hand predicts at small pressure a very high bulk modulus (~54 GPa) and shows the minimum in bulk modulus around 5-6 GPa. Similar trends are observed for the other elastic moduli (panels (b) and (c)): SHIK-1 gives a good prediction for the shear modulus and Young's modulus at zero pressure, whereas SHIK-2 underestimates these values by about 3 GPa and the BKS overestimates them significantly. For increasing pressure one find also in these moduli a minimum, the position and value of which are predicted well by the SHIK-1 potential, reasonably well by the SHIK-2 potential, but not so well for the BKS potential. For pressures beyond the minimum one finds that the elastic moduli for the SHIK-2 potential start to increase rapidly at 5-6 GPa, which is due to the onset of higher coordinated Si as seen in Fig. 8(b). In contrast to this SHIK-1 predicts a steady increase of the elastic moduli for pressures beyond the minimum, in agreement with the experimental observation. Finally we mention that all three potentials overestimate the Poisson's ratio given by $\nu = \frac{E}{2G} - 1$ and are not able to reproduce the distinct minimum seen in experiment, see panel (d). This failure is not that surprising since the Poisson's ratio is a very small quantity and hence the relative errors are large.

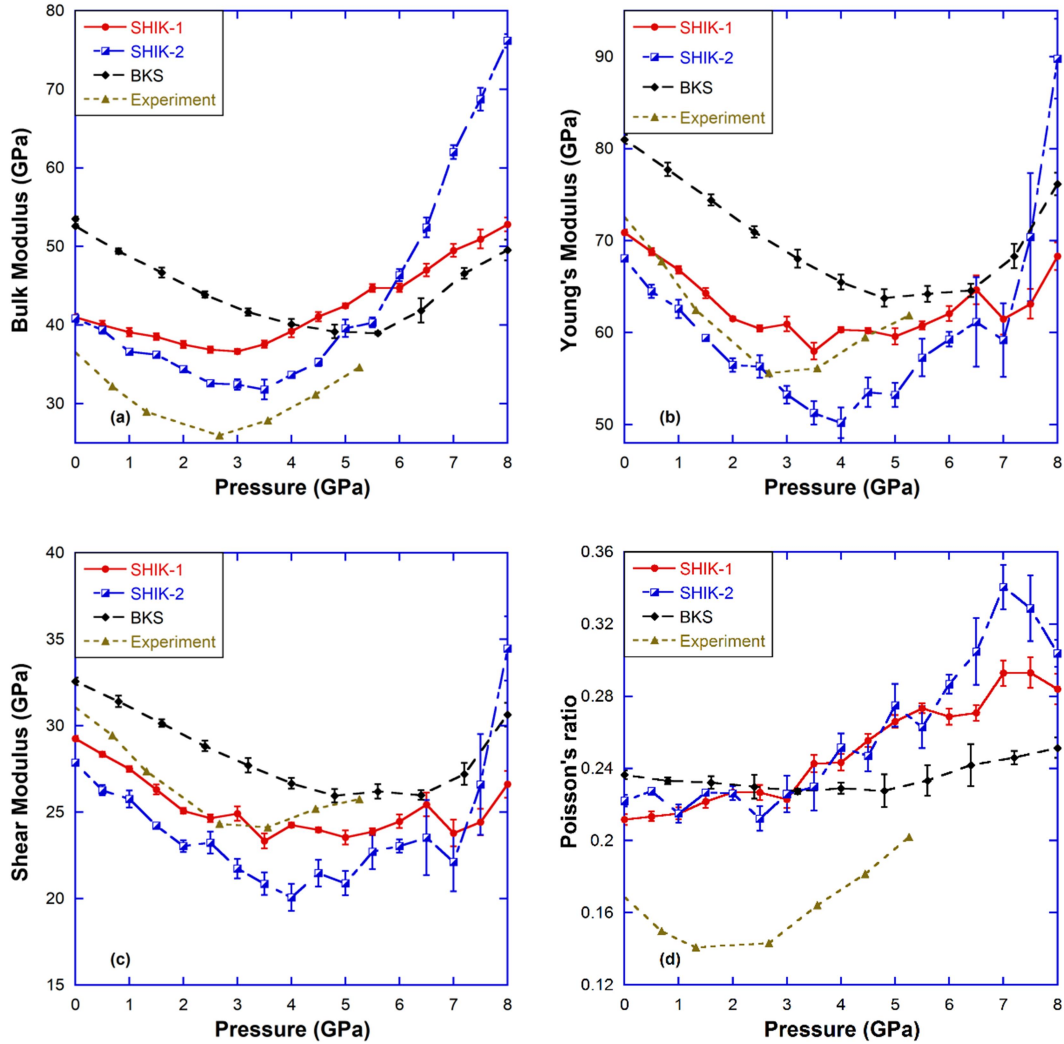

**Fig. 10** Variation of (a) bulk modulus, (b) Young's modulus, (c) shear modulus and (d) Poisson's ratio with pressure from the various potentials in comparison to experiment.[78]

## 4. Summary and Conclusions

In this work we have presented a new optimization scheme which can be used to obtain effective potentials. Although the results are for silica and we have chosen the Buckingham functional form, the approach is very general and hence can be easily applied to other atomic systems, such as multi-component glass-formers, as well.[79] By comparing the results for the SHIK-1 and SHIK-2 potentials, we have found that the choice of the cost function in conjunction with a higher short-range cutoff is crucial for obtaining a potential that is more reliable, especially in describing the resistance of the glass structure to compression. In the present case we have included in the cost function the pressure-dependence of the density and found that this improves significantly the capability of the pairwise potential to reproduce the response of structure and mechanical properties of silica glass to higher pressures.

In addition we have found that a good choice of the cost function also helps to remove inherent degeneracies in the parameter space due to the presence of various local minima in the cost function, i.e., certain local minima are removed. But even for the rather complex cost function that we have used here these degeneracies still exist. Therefore it is important to ensure that one has found a *good* local minimum instead of having been trapped in a high minimum. This can be achieved by starting the optimization from different starting points in the parameter space or by allowing upwards jumps with a certain probability to shake the system out of a local minimum. It should be noted that as with any optimization of this nature, this local minimum may not be the global minimum.

For many of the properties like the VDOS or mechanical properties, our previous study showed that the interaction among atoms plays a larger role than the atomic structure[35]. Hence using a glass structure quenched by one potential of the same functional form to measure some of these properties at room temperature and include them in the cost function can indeed help improve the interaction without the need for quenching during the optimization. We also confirmed that the quenched structure with optimized parameters was close to the one we used during the optimization process. In other words, glass structure prepared from different potentials of the same functional form can be very similar, but their properties are very different due to the different interaction among atoms.

Including only the high temperature structure in the cost function did not seem to improve the mechanical properties at room temperature[16]. We also tested whether using structure at multiple high temperatures in the cost function to train the potential parameters allowed to improve the glass properties and found this not to be the case. We believe that the large discrepancy in the mechanical properties is rather due to the interactions among the atoms rather than the differences in the glass structure. Including only structure and properties of high temperature liquid in the cost function is not sufficient to obtain interaction potentials that can reliably describe the glass, as at high temperature we are probing a different part of the potential energy landscape with no information about the low temperature region. Hence, to improve the interaction and predict better mechanical properties of glass, we include room temperature properties of glass in the cost function.

Using this new approach we have obtained a new potential, SHIK-1, which turns out to be not only clearly better than the most commonly used potential based on same functional form for silica glass, i.e., the BKS potential, but even more reliable than many three-body potentials. In particular this potential is more accurate in predicting the high temperature structure of silica liquid, the quench curves with improved fictive temperature at a given cooling rate, and the density and VDOS of silica glass at ambient conditions. The potential can also predict quite accurately the various elastic moduli and their response to pressure. The simplicity of the functional form also makes this potential computationally more efficient than the complicated three-body potentials and hence more suitable for simulations involving large length and/or time scales.

Although the SHIK-1 potential has the same functional form as the BKS potential, we have shown here that the former is clearly superior to the latter. Information contained in one $Si(OH)_4$ cluster, to which the BKS potential was optimized towards, is only within one tetrahedron. No information about the connectivity of tetrahedral units was included and thus less likely to be able to capture the structure of silica, like the radial distribution functions used in our study. One of the main differences between the new potential and the BKS potential is the reduction in the effective charge for Si from 2.4e to around 1.7755e (charge for O changes accordingly to maintain the charge neutrality). This reduction in the effective charges greatly improves the structure of silica liquid at high temperatures and the VDOS of silica glass at low temperatures. At the same time, the reduced charge makes the structure of silica glass less resistant to compression, as seen from the increase of the average coordination of Si around 5-6 GPa predicted by the SHIK-2 potential in Fig. 8(b). This in turn gives rise to the rapid increase in density and elastic moduli with increasing pressure. Including information on the density at various pressures in the cost function during the optimization, together with a higher short-range cutoff, shifts the onset of higher coordinated Si to higher pressures as seen for SHIK-1, which improves the density and elastic moduli at higher pressures. It is important to note that charge is just one of the fitting parameters and its effects may be compensated by others. A major challenge in developing interaction potentials is that the *effective* potential seen by a particle is the sum of many terms in the *bare* interaction potential, and hence it is very hard to separate out which parameters control which properties the most. To make this point clear, Fig. 11 below shows the pair interactions in the SHIK-1, SHIK-2, CHIK and BKS potentials. SHIK-1 and SHIK-2 potentials are very similar to each other but quite different from the BKS potential. The CHIK potential also resembles more of the new potentials than the BKS potential. It is interesting to note that two potentials like SHIK-1 and BKS can have very different pair interactions, but give a very similar structure. On the other hand, two potentials like SHIK-1 and SHIK-2 can have very similar pair interactions, but give very different properties.

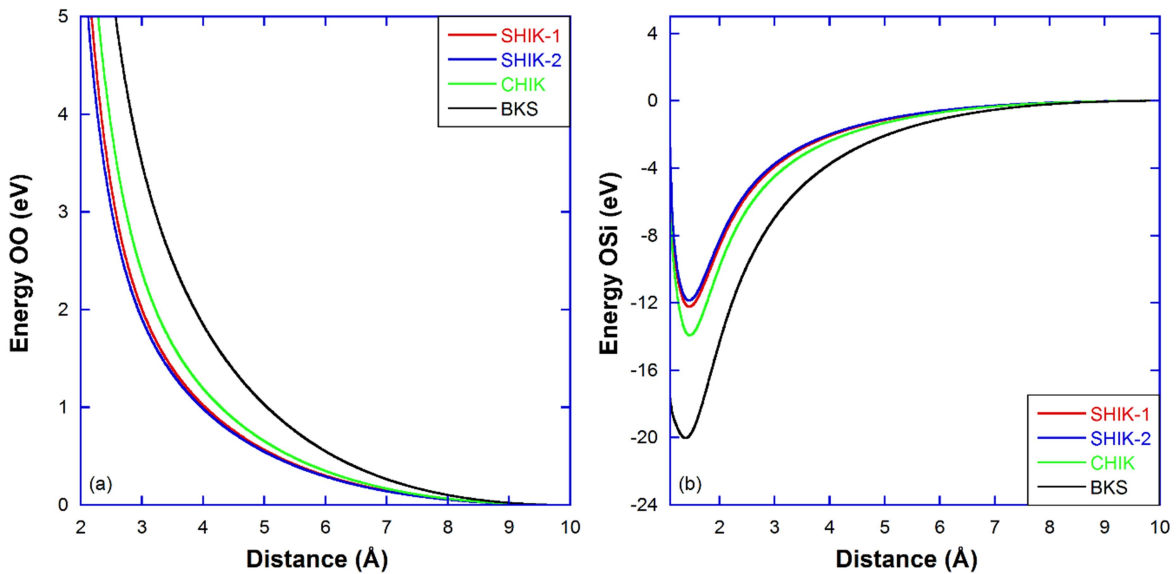

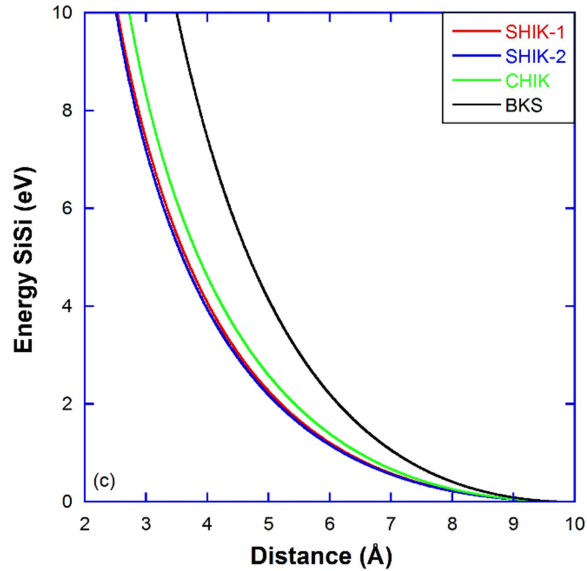

**Fig. 11** (a) O-O, (b) O-Si and (c) Si-Si pair interactions for SHIK-1, SHIK-2, CHIK and BKS potentials.

It is also important to note that the room temperature structures, as characterized by the RDFs and structure factors, are very similar for the various potentials, but that the elastic and vibrational properties are very different. This observation can be rationalized by the following facts: i) Since the RDFs and structure factors project the 3D structural information into 1D, details on the spatial distribution of atoms except distance are lost and thus they cannot differentiate one potential model from another; ii) Although structure models from different potentials can be very similar, the details in the interaction potential plays a major role in determining the elastic and vibrational properties. For either case, it necessitates the inclusion of additional constraints besides the 1D structural information from RDFs or structure factors into the cost function for optimizing potentials for MD simulations, and hence our approach to include the pressure dependence of the density is one possibility to provide this information. Our new optimization scheme demonstrates that reliable pairwise potentials based on the Buckingham functional form can be obtained for silica glass by adding VDOS and density besides RDFs into the cost function. As VDOS depends on not only the different features of the interaction potential as a whole (e.g., position, depth and curvature of the local potential well), but also on the 3-D atomic structure in the short and intermediate range, including it in the cost function helps to remove inherent degeneracies in the parameter space, thus improve other properties such as density or stiffness. Even though the overall features of the VDOS in this work did improve considerably over existing potentials, there are still some deficiencies, which are probably due to the intrinsic limitations of the simple Buckingham functional form or the effective potential in general. Including the mechanical properties in the cost function instead seems a promising approach for future optimizations. We included the structure at two temperatures with the aim of training the potential to predict the structure change more reliably with temperature. It is unclear whether this was indeed necessary and resulted in an improvement in the potential or data from more temperatures is required. Systematic studies including data from several temperatures and pressures with approaches such as machine learning[80–83] may indeed result in more reliable potentials.

**Acknowledgements**

L. Huang acknowledges the financial support from the US National Science Foundation under grant No. DMR-1105238 and DMR-1255378, and a Corning-CFES seed fund, as well computational resources from the Center for Computational Innovations at RPI. S. Ispas acknowledges HPC resources from GENCI (Grant No. x2016095045). S. Sundararaman acknowledges an NSF-IMI travel grant from DMR-0844014 to France to initiate this collaborative work.